# Enhancement of the Anti-Damping Spin Torque Efficacy of Platinum by Interface Modification

Minh-Hai Nguyen[1], Chi-Feng Pai (白奇峰)[1,†], Kayla X. Nguyen[1], David A. Muller[1,2], D. C. Ralph[1,2], R. A. Buhrman[1*]

[1]*Cornell University, Ithaca, New York 14853, USA*

[2]*Kavli Institute at Cornell, Ithaca, New York 14853, USA*

## Abstract

We report a strong enhancement of the efficacy of the spin Hall effect (SHE) of Pt for exerting anti-damping spin torque on an adjacent ferromagnetic layer by the insertion of $\approx 0.5$ nm layer of Hf between a Pt film and a thin, $\leq 2$ nm, $Fe_{60}Co_{20}B_{20}$ ferromagnetic layer. This enhancement is quantified by measurement of the switching current density when the ferromagnetic layer is the free electrode in a magnetic tunnel junction. The results are explained as the suppression of spin pumping through a substantial decrease in the effective spin-mixing conductance of the interface, but without a concomitant reduction of the ferromagnet's absorption of the SHE generated spin current.

* Author to whom correspondence should be addressed. Electronic mail: buhrman@cornell.edu

†*Current affiliation: Massachusetts Institute of Technology, Cambridge, Massachusetts 02139, USA*

The experimental determination that a current density $J_e$ flowing through certain high-atomic-number metals can generate a quite substantial transverse spin current density $J_s$ through the spin Hall effect (SHE)[1–3] has been a major factor in the recent focus on the study of spin-orbit interaction effects in heavy metal - ferromagnet (HM|FM) thin film multilayer systems. The fraction of this spin current that is absorbed by the ferromagnetic film generates a spin transfer torque on the FM, characterized by a spin-torque efficiency $\xi_{SH} \equiv (2e/\text{h}) J_s^{absorbed} / J_e \leq \theta_{SH} \equiv (2e/\text{h}) J_s / J_e$, where $\theta_{SH}$ is the "internal" spin Hall angle. For anti-damping spin-torque (ST) excitation, the mechanism by which the spin Hall torque can achieve magnetic manipulation using the least possible current, the critical current density scales $\propto \alpha / \xi_{SH}$ and the write energy $\propto (\alpha / \xi_{SH})^2 \rho$, where $\alpha$ is the Gilbert damping of the FM|HM bilayer and $\rho$ is the electrical resistivity of the HM. Large ST efficiencies have been measured for Pt, beta-phase or amorphous Ta, and beta-phase W films: $\xi_{SH}^{Pt} = 0.04 - 0.09$ [4–6], $\xi_{SH}^{\beta-Ta} \approx 0.12$ [7] and $\xi_{SH}^{\beta-W} \approx 0.3$ [8]. The large values of $\xi_{SH}$ for $\beta$-Ta and $\beta$-W, together with the relatively small values of damping for thin $\beta$-Ta and $\beta$-W|FM bilayers has enabled low-current ST switching and ST microwave excitation of the free electrode of nanoscale magnetic tunnel junctions (MTJs)[8], demonstrating the feasibility of the SHE for three-terminal memory device and ST nano-oscillator applications[9–11] as well as new classes of spin logic circuits[12–17]. However, the high resistivity of $\beta$-Ta and $\beta$-W, $\geq 180 \, \mu\Omega \cdot \text{cm}$, can be problematic when the write energy and device heating are important considerations. While the lower resistivity of Pt films, $\rho_{Pt} \approx 20 \, \mu\Omega \cdot \text{cm}$ for isolated films (which can be different from the averaged resistivity of thin Pt having adjoining metal layers of high resistivity - see Supplementary Material (SM) at [URL]), makes Pt seemingly more attractive for energy-efficient ST devices, the smaller value of $\xi_{SH}$ for Pt and a much higher damping for FM|Pt bilayers[4,7] greatly diminishes the effectiveness of anti-damping ST for Pt devices.

Here we report that a thin, $\approx 0.5$ nm, Hf layer inserted between a Pt film and a thin $Fe_{60}Co_{20}B_{20}$ (FeCoB) layer causes large reductions in both the current density and write energy needed for ST switching. The presence of the Hf reduces the Gilbert damping $\alpha$ by more than a factor of two by suppressing spin pumping, and at the same time results in $\xi_{SH}^{Pt|Hf} \approx 0.12$, approximately 2 times higher than the spin torque efficiency reported with Pt|$Ni_{81}Fe_{19}$ bilayers. Pt|Hf|FeCoB is therefore a preferred SHE structure for use in anti-damping ST applications. Our work suggests that there may be additional opportunities for the enhancement of spin Hall torque effects through the further optimized modification of HM|FM interfaces.

Understanding the consequences of the Hf insertion layer requires an analysis of the processes contributing to magnetic damping and spin transmission at an HM|FM interface. The phenomenon of spin pumping, which is typically analyzed via use of the drift-diffusion equation,[18] increases the magnetic damping $\alpha$ in HM|FM structures compared to $\alpha_0$, the intrinsic damping parameter in the absence of the HM, because the precession of the FM magnetization leads to a loss of spin angular momentum in the HM[19] resulting in:

$$\alpha = \alpha_0 + \frac{\gamma \hbar^2}{2e^2 M_s t_{FM}} G_{eff}^{\uparrow\downarrow}. \qquad (1)$$

Here $\gamma = 1.76 \times 10^{11}$ HzT$^{-1}$ is the gyromagnetic ratio, $M_s$ is the saturation magnetization of the FM, $t_{FM}$ is the thickness of the FM layer, and $G_{eff}^{\uparrow\downarrow}$ is the "effective spin-mixing conductance" of the HM|FM interface. $G_{eff}^{\uparrow\downarrow}$ can be expressed in terms of the bare spin mixing conductance of the

interface $G^{\uparrow\downarrow}$ (here we are assuming that $|\text{Re}\,G^{\uparrow\downarrow}| \gg |\text{Im}\,G^{\uparrow\downarrow}|$[20]), and the spin conductance of the HM layer, $G_{\text{ext}} \equiv \tanh(t_{\text{HM}}/\lambda_s)/(2\rho_{\text{HM}}\lambda_s)$, as[20–23]

$$G_{\text{eff}}^{\uparrow\downarrow} = \frac{G^{\uparrow\downarrow}}{1 + G^{\uparrow\downarrow}/G_{\text{ext}}}. \qquad (2)$$

Pt has a relatively large value of $G_{\text{eff}}^{\uparrow\downarrow}$ and is therefore a good "spin sink", since the typical resistivity of Pt films, $\rho_{\text{Pt}} \approx 20-25\ \mu\Omega\cdot\text{cm}$ in combination with a spin attenuation length $\lambda_s^{\text{Pt}} \approx 1.2-1.4$ nm [24,25] results in $G_{\text{ext}} \approx 1.8\times10^{15}\ \Omega^{-1}\text{m}^{-2}$ (assuming $t_{\text{HM}} \gg \lambda_s$), while the bare mixing conductance of common Pt|FM interfaces is usually of a similar value, *e.g.* $G^{\uparrow\downarrow} \approx 1.2\times10^{15}\ \Omega^{-1}\text{m}^{-2}$ has been reported[25] for Pt|Py (Py=Ni$_{81}$Fe$_{19}$). Consequently, $G_{\text{eff}}^{\uparrow\downarrow}$ for Pt|FM bilayers is considerably higher, $\geq 0.7\times10^{15}\ \Omega^{-1}\text{m}^{-2}$, than that found, for example, for the *β*-W|FeCoB system,[26] $0.16\times10^{15}\ \Omega^{-1}\text{m}^{-2}$. When $t_{\text{FM}}$ is small, as it must be in ST devices, the large value of $G_{\text{eff}}^{\uparrow\downarrow}$ causes a large increase in $\alpha$ above $\alpha_0$ for Pt|FM bilayers (*e.g.*, more than a factor of 4 for a 1.8 nm FeCoB layer, see below). This greatly reduces the efficacy of Pt for anti-damping ST applications, although the SHE of Pt is still quite effective for driving the displacement of domain walls in perpendicularly magnetized free layers where the increased damping is not an issue[27,28].

Turning to the spin torque efficiency $\xi_{\text{SH}}$, this is of course affected by the net interfacial spin transmissivity in the opposite direction, that is by the extent to which spin currents generated by the SHE in the HM are transmitted through the HM|FM interface to exert a spin torque on the FM. The drift-diffusion analysis for this situation[20,21] indicates that unless $\text{Re}\,G^{\uparrow\downarrow} \gg G_{\text{ext}}$ there will be substantial spin back-flow from the interface, which reduces the efficiency of the SHE (relative to the internal spin Hall angle, $\theta_{SH}$) in exerting a damping-like spin torque on the FM:

$$\xi_{SH} = \theta_{SH} \times \frac{G^{\uparrow\downarrow} \tanh\left(\frac{t_{HM}}{2\lambda_s}\right) \tanh\left(\frac{t_{HM}}{\lambda_s}\right)}{G^{\uparrow\downarrow} + G_{ext}} = \theta_{SH} \times \frac{G^{\uparrow\downarrow}_{eff}}{G_{ext}} \tanh\left(\frac{t_{HM}}{2\lambda_s}\right) \tanh\left(\frac{t_{HM}}{\lambda_s}\right). \quad (3)$$

This reduction can be quite significant. For example, applying the analysis above to the Pt|Py interface yields $\xi_{SH} \approx 0.25\, \theta_{SH}^{Pt}$ indicating that the lower bound of $\xi_{SH} \approx 0.05$ as established by the ST ferromagnetic resonance (ST-FMR) study of Pt|Py bilayers of Liu et al.[4,24] is considerably lower than the actual internal spin Hall angle of the Pt film $\theta_{SH}^{Pt} \approx 0.20$[29]. This is similar to the result of the same analysis applied to Pt|Co and Pt|CoFe interfaces[29,30]. This analysis suggests that by an appropriate choice of materials and control of the interface structure, we could possibly achieve higher ST efficiencies, and indeed recent studies using different Pt|FM combinations have reported[29,30] $\xi_{SH} \approx 0.1$ in some cases.

With the objective of investigating means to suppress spin-pumping and to enhance, or at least not degrade, the spin torque efficiency of Pt for MTJ switching applications, we produced ||Ta(1)|Pt(4)|Hf($t_{Hf}$)|FeCoB($t_{FeCoB}$)|MgO(1.6)|Ru(2) and ||Ta(1)|Pt(4)|Hf($t_{Hf}$)|FeCoB($t_{FeCoB}$)|MgO(1.6)|FeCoB(4)|Hf(5)|Ru(5) multilayer films (Here || represents the thermally-oxidized Si substrate and the numbers in parentheses are thicknesses in nm). (See SM) The high resistivity Ta was used for adhesion and smoothing purposes, while Hf was chosen for this investigation because initial ferromagnetic resonance studies (FMR) studies indicated a low $G^{\uparrow\downarrow}_{eff}$ for Hf|FeCoB and previous work has demonstrated that there are negligible current-induced spin-orbit torques produced at Hf|FM interfaces[31]. The Hf thickness $t_{Hf}$ was

varied in fine steps from 0.33 to 0.76 nm, while the FeCoB layer thicknesses $t_{FeCoB}$ were 1.6 and 1.8 nm. We also fabricated and measured control samples with $t_{Hf} = 0$ (i.e., no Hf spacer). The samples were annealed at 300 C for 30 minutes in a background pressure $< 10^{-7}$ Torr.

Since certain transition metal elements when incorporated into magnetic tunnel junction structures can be quite mobile, either during deposition and subsequent annealing steps, we investigated this possibility with respect to the Hf insertion layer by using electron energy loss spectroscopy (EELS)[32] to study the spatial-dependent composition of some of our samples in a 100 keV Nion UltraSTEM. Fig. 1 shows the EELS data for a $t_{FeCoB} = 1.6$ nm, $t_{Hf} = 0.5$ nm sample. Since the obtained intensity is proportional to the relative distribution of the element, it is readily seen that a portion of Hf has diffused through the FeCoB layer into the MgO layer, where it is now oxidized. By integrating over the intensity, the amount of Hf in between the Pt and FeCoB layers is estimated to be ~70% of the total amount of Hf deposited which corresponds to a thickness of ~0.35 nm for this nominal 0.5 nm Hf sample. This indicates that a thin, conformal and continuous Hf spacer of approximately two atomic layers or so in thickness is formed on the surface of Pt layer, which is consistent with the high (negative) formation enthalpy of HfPt compounds[33,34].

The magnetic properties of the FeCoB layer in the first set of multilayers were characterized by SQUID magnetometry and anomalous Hall measurements (see SM) which indicated a saturation magnetization $M_s = (1.56 \pm 0.06) \times 10^6$ A/m, and also an apparent "magnetic dead layer" thickness $t_d = 0.7 \pm 0.1$ nm. By fitting the data from measurement of the effective magnetic anisotropy energy $K_{eff} t_{FeCoB}^{eff}$ as a function of FeCoB effective thickness $t_{FeCoB}^{eff} = t_{FeCoB} - t_d$ to the standard model for the thickness dependence of the magnetic anisotropy[35]

$$K_{eff} t_{FeCoB}^{eff} = \left(K_V - (1/2)\mu_0 M_s^2\right) t_{FeCoB}^{eff} + K_S \qquad (4)$$

the interface and bulk anisotropy energy densities are estimated to be $K_S = 0.45 \pm 0.03$ mJ/m$^2$ and $K_V = 0.60 \pm 0.03$ MJ/m$^3$, respectively. This value of $K_S$ is smaller than typical for Ta|FeCoB|MgO multilayers, while $K_V$ is similar to a recent report[36].

Finally, measurements of the in-plane effective demagnetization field $\mu_0 M_{eff}$ for $t_{FeCoB} = 1.6$ nm and $t_{FeCoB} = 1.8$ nm ||Ta(1)|Pt(4)|Hf($t_{Hf}$)|FeCoB($t_{FeCoB}$)|MgO(1.6)|Ru(2) samples indicated that the insertion of a thin layer Hf at the interface of Pt and FeCoB has a significant effect on $\mu_0 M_{eff}$, with a local *minimum* at $t_{Hf} = 0.5$ nm for both series (see SM). We tentatively attribute this behavior to the role of the Hf insertion layer in both reducing the positive volume anisotropy effect from elastic strain from the underlying Pt, and in enhancing the surface anisotropy energy through reduction of strain at the FeCoB|MgO interface.

In Fig. 2(a) we show $\alpha(t_{Hf})$, determined by frequency-dependent ST-FMR measurements[4] (see SM), for the two different FeCoB thicknesses, 1.6 nm and 1.8 nm with the results clearly demonstrating that a deposited Hf layer as thin as 0.35 nm, or even less, is effective in greatly reducing the spin-pumping-induced increase in $\alpha$. All of the samples with the Hf insertion layer exhibit a decrease in $\alpha$ by a factor of two or more compared to Pt|FeCoB(1.8 nm) with no insertion layer (also shown in Fig. 2(a)). We quantified the effect of the 0.5 nm Hf insertion layer on $G_{eff}^{\uparrow\downarrow}$ of a series of $t_{Hf} = 0.5$ nm samples as the function of $t_{FeCoB}^{eff}$. Fig. 2(b) shows the best fit to the damping coefficient $\alpha(t_{FeCoB}^{eff})$ data (solid line in Fig. 2(b)) to equation (1) which yields $\alpha_0 \approx 0.006$ (broken line) and $G_{eff}^{\uparrow\downarrow} \approx 0.24 \times 10^{15}$ $\Omega^{-1}$m$^{-2}$. This $G_{eff}^{\uparrow\downarrow}$ value is nearly as low as the value $0.16 \times 10^{15}$ $\Omega^{-1}$m$^{-2}$ observed in the $\beta$-W|FeCoB system[26]. Similar measurements made on a series of $t_{Hf} = 0$ control samples (see SM) yielded

$G_{\text{eff}}^{\uparrow\downarrow} \approx (1.1 \pm 0.1) \times 10^{15} \, \Omega^{-1} \text{m}^{-2}$, similar, although somewhat higher, than the previous results for Pt|Ni$_{81}$Fe$_{19}$, confirming the strong effectiveness of the insertion of a nominal 0.5 nm Hf layer in suppressing spin pumping, as reflected by the large reduction of $\Delta\alpha = \alpha - \alpha_0$ shown in Fig. 2(a).

To determine $\xi_{\text{SH}}$ for the Pt|Hf|FeCoB trilayers we measured the ST switching current of a FeCoB free layer in a MTJ, which is the application for which we seek to improve the spin Hall efficacy[8]. To accomplish these switching current measurements, we patterned the second set of multilayers by electron beam lithography and ion milling (described in SM) into three terminal SHE-MTJ devices which consisted of elliptical FeCoB|MgO|FeCoB MTJs, typically with lateral dimensions $\approx 50 \times 180 \text{ nm}^2$, on top of a Ta|Pt|Hf microstrip approximately 1.2 μm wide as shown schematically in Fig. 3(a). The magnetization of the free FM layer could be controlled either by an in-plane external field along the major axis of the tunnel junction or a direct current through Pt layer, and the orientation of the magnetic free layer can be determined by the differential resistance of the MTJ.

Figure 3(b)-(d) show results for $t_{\text{FeCoB}} = 1.6 \text{ nm}$, $t_{\text{Hf}} = 0.5 \text{ nm}$ devices. From the field switching behavior the coercivity $\mu_0 H_c$ is determined to be about 4.5 mT and the tunneling magnetoresistance (TMR) is 80%. Figure 3(b) shows the current-switching behavior for a $50 \times 180 \text{ nm}^2$ MTJ, 1.2 μm channel device at a ramp rate 0.0013 mA/s, for which the switching occurs at average critical currents $I_c = \pm 0.4 \text{ mA}$. The switching currents at different ramp rates are shown in Fig. 3(c). By fitting the data to the thermally assisted spin torque switching model[37] we find that the zero-thermal-fluctuation switching current is $I_0 = 0.71 \pm 0.08 \text{ mA}$ which is, considering the geometry of the device and assuming that all current flows through the comparatively low-resistivity 4 nm Pt layer, equivalent to a current density of $J_0 = (1.5 \pm 0.2) \times 10^{11}$ A/m$^2$. The same measurement and analysis were also performed for $t_{FeCoB} = 1.6$ nm, $t_{Hf} = 0.5$ nm, $70 \times 240$ nm$^2$ devices with different channel widths. As shown in Fig. 3(d), we confirmed that the switching current $I_0$ varies linearly with the channel width $w$, as expected, and that the average zero-fluctuation switching current density $J_0 = (1.6 \pm 0.1) \times 10^{11}$ A/m$^2$ for that series of devices is consistent with that of the $50 \times 180$ nm$^2$ MTJ, 1.2 µm-wide channel device.

We used the results of the measurements of the switching current density $J_0$ as plotted in Fig. 4(a) as a function of $t_{Hf}$ to calculate $\xi_{SH}$, using the measured values for $M_{eff}$ and $\alpha$ mentioned above and the formula[8,38]

$$\xi_{SH} = \frac{2e}{\hbar} \mu_0 M_s t_{FeCoB}^{eff} \alpha \left( H_c + \frac{M_{eff}}{2} \right) / J_0 \qquad (5)$$

Those latter results are plotted in Fig. 4(b). For $0 \leq t_{Hf} \leq 0.6$ nm, the spin torque efficiency fluctuates about the average value $\xi_{SH} = 0.10$, with a peak value $\xi_{SH} = 0.12 \pm 0.02$ at both $t_{Hf} = 0.0$ nm and $t_{Hf} = 0.5$ nm. While a quantitative analysis using the drift-diffusion model of these results for the spin torque efficiency of the Pt|Hf|FeCoB trilayer structures is conceptually challenging in the $t_{Hf} \approx 0.5$ nm ultrathin limit, this is less of an obvious concern for of the Pt|FeCoB bilayer samples. If we use $\rho_{Pt} = 24$ µΩ·cm as determined for our samples (see SM) and $\lambda_s^{Pt} = 1.2$ nm [25] (determined for samples having the same electrical resistivity), we have that for the Pt layer $G_{ext} = 1.7 \times 10^{15}$ Ω$^{-1}$m$^{-2}$. Equations (2) and (3) then yield $G^{\uparrow\downarrow} = G_{ext} G_{eff}^{\uparrow\downarrow} / (G_{ext} - G_{eff}^{\uparrow\downarrow}) \approx 3.1 \times 10^{15}$ Ω$^{-1}$m$^{-2}$ and $\xi_{SH}^{Pt|FeCoB} / \theta_{SH}^{Pt} = 0.65$, where the latter is a considerably higher ratio than reported for a Pt|Py bilayer $\xi_{SH}^{Pt|Py} / \theta_{SH}^{Pt} = 0.25$ [29], which signifies that our Pt|FeCoB interface has a significantly higher spin current transmissivity. With

$\xi_{SH}^{Pt|FeCoB} / \theta_{SH}^{Pt} = 0.65$ the high spin torque efficiency $\xi_{SH}^{Pt|FeCoB} = 0.12 \pm 0.02$ obtained from the switching measurements indicates that $\theta_{SH}^{Pt} = 0.18 \pm 0.03$, quite consistent with the spin Hall angle values recently reported from analyzes of experiments on Pt|Py, Pt|Co and Pt|CoFe systems[29,30].

Returning to the results from the devices with the Hf insertion layer, while the substantial decrease in $\xi_{SH}$ when $t_{Hf}$ is increased from 0.6 nm to 0.76 nm is perhaps qualitatively consistent with an increased attenuation of the spin current by a thicker Hf layer[31], the lack of significant variation of $\xi_{SH}$ for $0 \leq t_{Hf} \leq 0.6$ nm is quite surprising in light of the strong suppression of spin pumping, which represents a factor of 4 reduction of $G_{eff}^{\uparrow\downarrow}$ by a Hf insertion only one to two atomic layers thick. Within the drift-diffusion analysis the most straightforward explanation for this spin pumping reduction is that the spin mixing conductance $G_{Hf|FeCoB}^{\uparrow\downarrow} \ll G_{Pt|FeCoB}^{\uparrow\downarrow}$, with the alternative being that the Hf layer has a very low spin conductance, $1/(2\rho_{Hf}\lambda_s^{Hf})$, together with $t_{Hf} \geq \lambda_s^{Hf}$. In light of the measured $\xi_{SH}(t_{Hf})$ results (Fig. 4(b)) there are fundamental challenges for both explanations. A low $G^{\uparrow\downarrow}$ will enhance the back flow of the spin current from the Hf|FeCoF interface, lowering $\xi_{SH}$ as implied by equation (3). This could be counteracted if $G_{ext}$ is also lowered by a similar degree by the Hf insertion, but since the experimental evidence is that Hf has no significant SHE a low $G_{ext, Hf}$ would also result in a strong attenuation of the spin current from the Pt before it reaches the Hf|FeCoB interface (see SM for further discussion). Another possible explanation could be that the Hf insertion results in a decreased $G_{ext, Pt}$, or an enhanced $\theta_{SH}^{Pt}$, through intermixing, but the similarity of the averaged resistivity of the Pt layer with and without the Hf insertion, together with results of experiments with PtHf alloys that will be reported elsewhere, appear to make this alternative explanation unlikely. Given these contradictions between the spin pumping and spin backflow (spin accumulation) predictions of the drift-diffusion equation and the results reported here, we tentatively conclude that with the very thin layers that are employed in this system, where interfacial scattering is a dominant factor, drift-diffusion simply does not provide an adequate understanding of the essential spin transport details. To achieve that, a Boltzmann equation analysis of the interfacial spin transmissivity and a more detailed description of the electronic structure of the interface is likely to be required.

In summary, we have maintained a high spin torque efficiency $\xi_{SH} = 0.12 \pm 0.02$ in Pt|FeCoB based three terminal SHE-MTJ devices, while substantially reducing the effect of spin pumping in increasing the damping of the thin FeCoB free layer. We have achieved this by introducing a thin, nominally 0.5 nm, Hf layer between the Pt SHE layer and the FeCoB. This reduced the magnetic damping to $\alpha \approx 0.012$ without significantly changing the spin torque efficiency and thus subsequently lowered the SHE switching current density to $\approx 1.6 \times 10^{11}$ A/m$^2$. This value is approximately a factor of 2 lower than achieved previously in similar Ta|Co$_{40}$Fe$_{40}$B$_{20}$ SHE-MTJ devices having much higher resistivity. The decrease in the damping can be attributed to a suppression of spin pumping brought about by a large reduction of the effective spin mixing conductance $G_{eff}^{\uparrow\downarrow}$ of Pt|Hf|FeCoB compared to Pt|FM but in a way that does not reduce the absorption of spin current at the FM interface. Although further theoretical investigation is necessary for a complete explanation and optimally to guide further improvements, the experimental determination that Pt|Hf|FeCoB samples can provide high spin torque efficiency together with an electrical resistivity much less than for $\beta$-Ta and $\beta$-W, demonstrates clearly that Pt|Hf provides an attractive alternative to those materials for anti-damping SHE torque logic devices for which impedance and low excitation power are important criteria.

**ACKNOWLEDGMENTS**

The authors thank J. Park, Y. Ou of Cornell University; C. T. Boone of NIST; and L. H. V. Leão of Federal University of Pernambuco (Brazil) for fruitful discussions. We thank S. Parkin of IBM Almaden Research Center for sharing a manuscript reporting a related study prior to publication. This work was supported in part by the NSF/MRSEC program (DMR-1120296) through the Cornell Center for Materials Research, ONR, and the Samsung Electronics Corporation. We also acknowledge support from the NSF (Grant ECCS-0335765) through use of the Cornell Nanofabrication Facility/National Nanofabrication Infrastructure Network. A patent application has been filed on behalf of the authors regarding technology applications of some of the findings reported here.

**FIGURES**

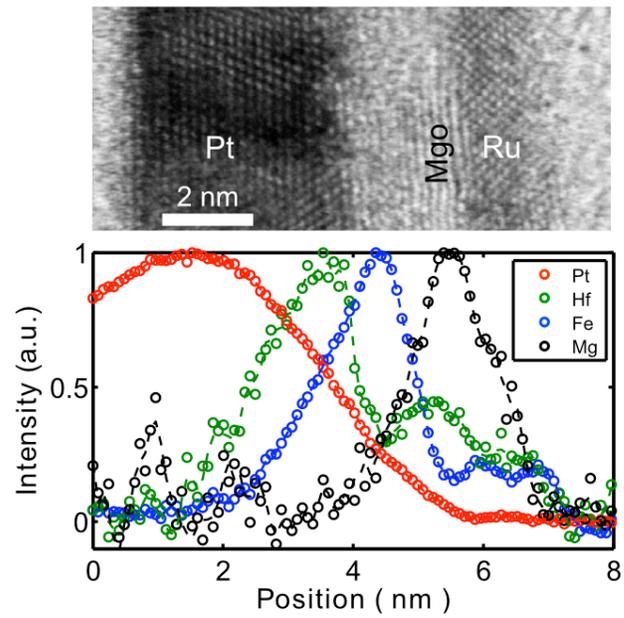

**Figure 1**. ABF-STEM (annular bright field - scanning transmission electron microscopy) image of ||Ta(1)|Pt(4)|Hf(0.5)|FeCoB(1.6)|MgO(1.6)|Ru(2) sample and the corresponding EELS line profile that shows Hf diffusion through the FeCoB into the MgO.

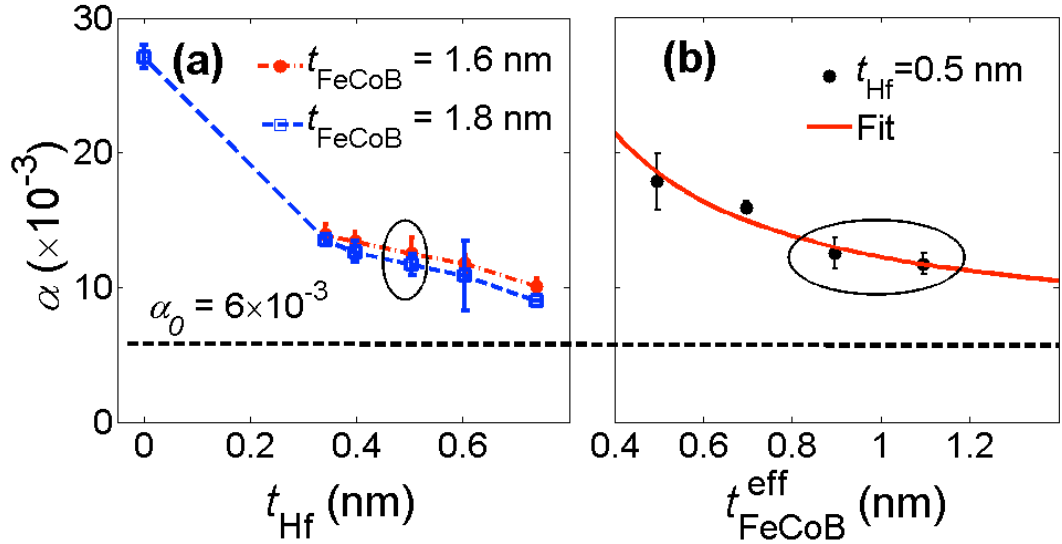

**Figure 2**. Gilbert damping parameter α of $\|Ta(1)|Pt(4)|Hf(t_{Hf})|FeCoB(t_{FeCoB})|MgO(1.6)$ samples measured by frequency-dependent ST-FMR. (**a**) Damping parameter versus Hf thickness $t_{Hf}$ for the $t_{FeCoB}=1.6$ nm (circles) and $t_{FeCoB}=1.8$ nm (squares) samples. The horizontal broken line indicates the fitted damping parameter (0.006) for an isolated FeCoB layer. (**b**) Damping parameter versus FeCoB effective thickness of the $t_{Hf}=0.5$ nm samples. The solid line shows the fitting result from which the magnetic damping parameter of isolated FeCoB film is estimated. The ellipses in (**a**) and (**b**) indicate the same data points.

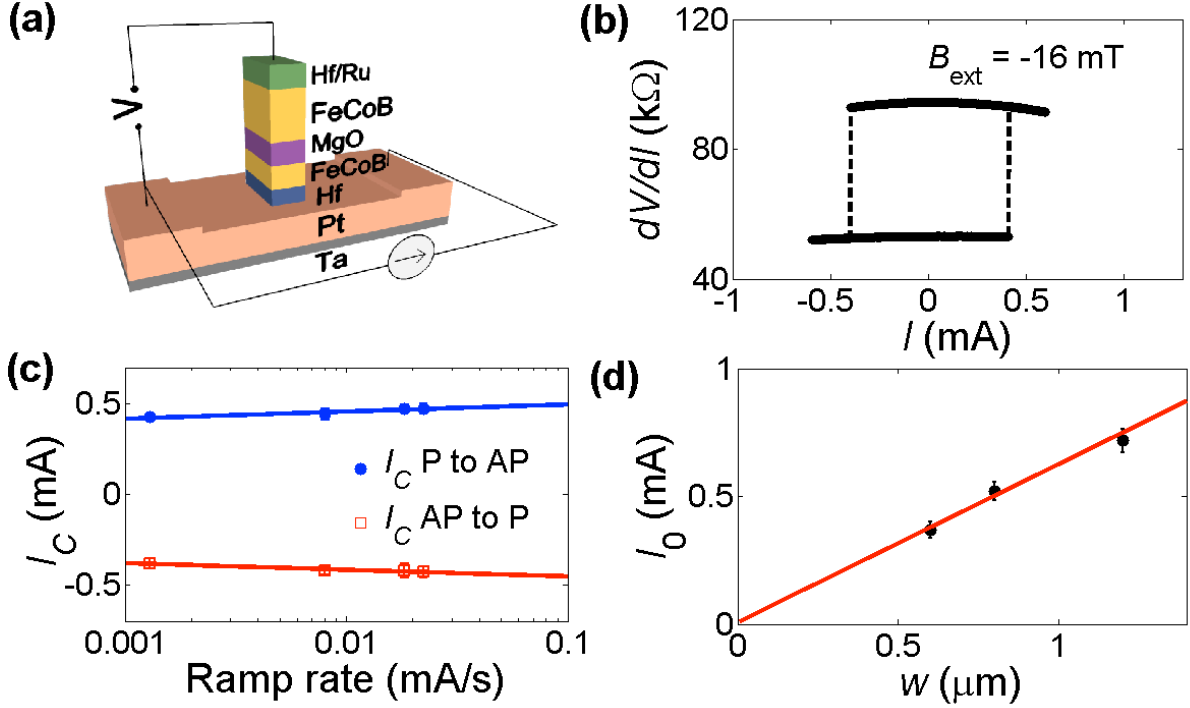

**Figure 3.** Current-induced switching behavior of ||Ta(1)|Pt(4)|Hf(0.5)|FeCoB(1.6)|MgO(1.6)|FeCoB(4) three-terminal devices. **(a)** Schematic structure of ||Ta|Pt|Hf|FeCoB|MgO|FeCoB three-terminal SHE-MTJ devices. **(b)** Differential resistance versus total current $I$ applied to the channel at a ramp rate 0.0013 mA/s for $50 \times 180$ nm$^2$ MTJ with a 1.2 μm channel. The switching currents are determined to be $I_c \approx \pm 0.4$ mA. Broken lines connect the data points, indicating the magnetic switching events. **(c)** Plot of switching currents at different ramp rates of 0.0013 mA/s for a $50 \times 180$ nm$^2$ MTJ with a 1.2 μm channel. Solid lines show fitted results. **(d)** $I_0$ versus channel width $w$ of $70 \times 240$ nm$^2$ devices. The linear fit (line) gives the average current density $J_0 = (1.55 \pm 0.12) \times 10^{11}$ A/m$^2$.

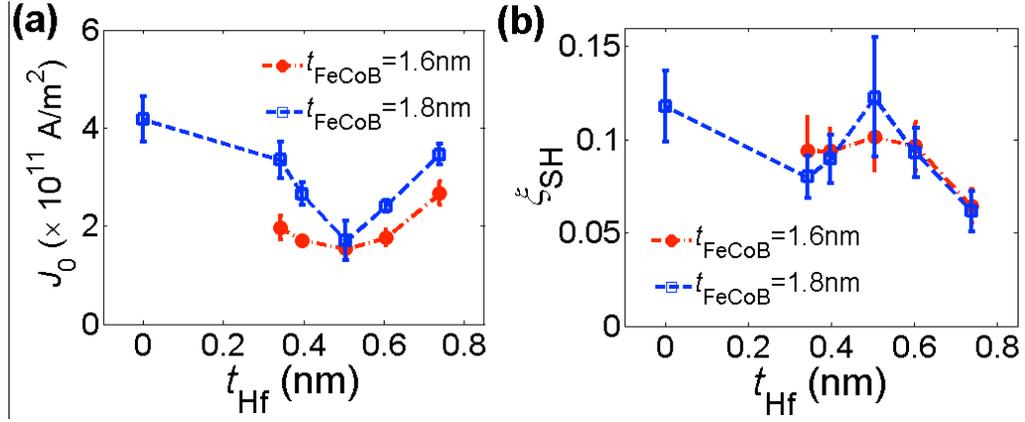

**Figure 4.** (a) Switching current density $J_0$ and (b) spin Hall torque efficiency $\xi_{SH}$ versus Hf thickness for $t_{FeCoB} = 1.6$ nm (circles) and $t_{FeCoB} = 1.8$ nm (squares) samples. $J_0$ achieves a minimum at $t_{Hf} = 0.5$ nm. Within the uncertainty, $\xi_{SH} \approx 0.10$ for $t_{Hf} < 0.6$ nm with local maxima $\approx 0.12$ at $t_{Hf} = 0$ and 0.5 nm, but then decreases for thicker Hf spacer. Dashed lines connect the data points to guide the eye.

SUPPLEMENTARY MATERIAL

# Enhancement of the Anti-Damping Spin Torque Efficacy of Platinum by Interface Modification

Minh-Hai Nguyen[1], Chi-Feng Pai[1†], Kayla X. Nguyen[1], David A. Muller[1,2], D. C. Ralph[1,2] and R.A. Buhrman[1*]

[1]Cornell University, Ithaca, New York 14853, USA

[2]Kavli Institute at Cornell, Ithaca, New York 14853, USA

[†]Current affiliation: Massachusetts Institute of Technology, Cambridge, Massachusetts 02139, USA

[*]rab8@cornell.edu

## Contents

**S1. Sample Preparation**

The multilayer films were produced by DC sputtering (radio frequency sputtering for the MgO layer) from 2-inch planar magnetron sources onto thermally-oxidized Si substrates in a sputter system with a base pressure $<4\times10^{-8}$ Torr. The target to substrate separation was approximately 18 cm. This separation together with an oblique orientation of the target to the substrate resulted in a low deposition rate of $\approx 0.01$ nm/s (Pt: 0.017 nm/s, Hf: 0.021 nm/s, FeCoB: 0.0077 nm/s) with DC sputtering conditions of 2 mTorr Ar and 30 watts power. The multilayer stacks ||Ta(1)|Pt(4)|Hf($t_{Hf}$)|FeCoB($t_{FeCoB}$)|MgO(1.6)|Ru(2) that were used for ST-FMR, anomalous Hall, and SQUID magnetometry measurements were patterned into $10\times20$ μm$^2$ microstrips by photolithography. The stacks ||Ta(1)|Pt(4)|Hf($t_{Hf}$)|FeCoB($t_{FeCoB}$)|MgO(1.6)|FeCoB(4)|Hf(5)|Ru(5) that were used for current-induced switching experiments were patterned into 3-terminal SHE-MTJ devices[1,2] which consisted of an elliptical FeCoB|MgO|FeCoB MTJ of typical size $\approx 50\times180$ nm$^2$ on top of a Hf|Pt|Ta channel of width $0.6-1.2$ μm (see Fig. 3**a**) by electron-beam lithography. The films were etched by an ion mill equipped with a mass spectroscopy system for endpoint detection. The samples were annealed at 300 C for 30 minutes in a vacuum tube furnace with a background pressure $<10^{-7}$ Torr.

## S2. Measurements

The damping parameters were measured by the frequency-dependent spin torque ferromagnetic resonance[3] (ST-FMR) technique in which an external magnetic field was applied in-plane at a 45° angle with respect to the 10×20 μm² microstrip. A radio frequency signal of power 12 dBm and frequency 5-10 GHz was applied to the microstrip and the DC output signal was detected through a bias-tee by a lock-in amplifier. From the frequency $f$ dependence of the linewidth $\Delta$, the damping coefficient was calculated from the linear fit as $\alpha = (\gamma/2\pi)\mathrm{d}\Delta/\mathrm{d}f$ (Fig. S1(a)) where $\gamma = 1.76 \times 10^{11}$ Hz·T$^{-1}$ is the gyromagnetic ratio.

The demagnetization fields were determined by anomalous Hall measurement with an applied magnetic field swept up to 1.5 T perpendicular to the sample plane as shown in Fig. S1(b). The film magnetizations were measured by SQUID magnetometry with an in-plane magnetic field.

The current-switching measurement on the three-terminal devices was performed by applying an external magnetic field from a Helmholtz electric magnet parallel to the major axis of the nanopillar to null out the average dipole field on the thin free layer from the thicker, fixed layer. A swept direct current was applied through the Pt channel. The tunnel junction was connected in series with a 10 MΩ resistor and a lock-in amplifier was used to measure its total differential resistance.

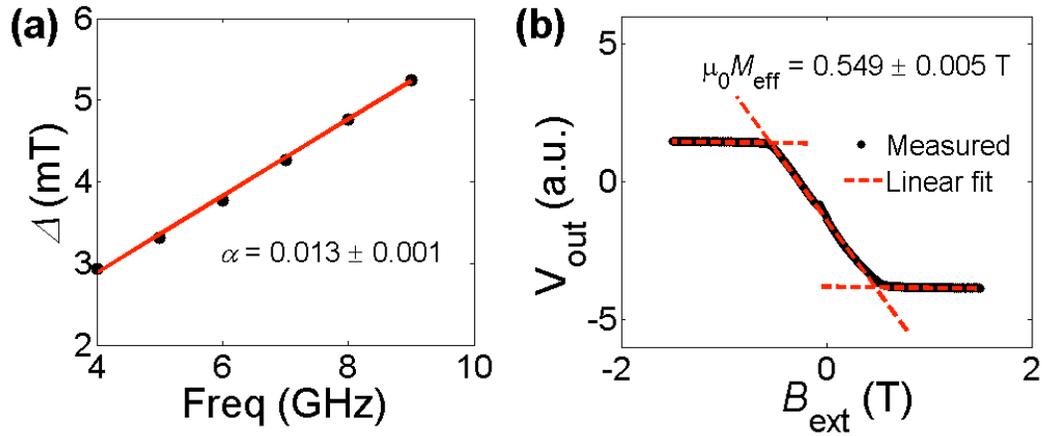

**Figure S1. (a)** Determination of Gilbert magnetic damping coefficient of a ||Ta(1)|Pt(4)|Hf(0.5)|FeCoB(1.6)|MgO(1.6)|Ru(2) sample by frequency-dependent ST-FMR measurement. **(b)** Determination of the effective demagnetization field of the same sample by anomalous Hall measurement with a perpendicularly applied magnetic field $B_{EXT}$. Lines show fitted results.

## S3. Magnetic properties of Pt|Hf|FeCoB structure

The magnetic properties of the FeCoB layer in the first set of multilayers were characterized by SQUID magnetometry and anomalous Hall measurements (see S2). In Fig. S2(a) we show the thickness dependence of the FeCoB magnetic moment (per unit area). The linear fit indicates that the saturation magnetization $M_s = (1.56 \pm 0.06) \times 10^6$ A/m, and also that the FM has an apparent "magnetic dead layer" thickness $t_d = 0.7 \pm 0.1$ nm. In Fig. S2(b) we plot the effective magnetic anisotropy energy $K_{\text{eff}} t_{\text{FeCoB}}^{\text{eff}}$ as a function of $t_{\text{FeCoB}}^{\text{eff}}$ for $t_{\text{Hf}} = 0.5$ nm, where $t_{\text{FeCoB}}^{\text{eff}} = t_{\text{FeCoB}} - t_d$ is the FeCoB effective thickness. Below $t_{\text{FeCoB}}^{\text{eff}} = 0.5$ nm, the magnetic anisotropy transitions from in-plane to out-of-plane. By fitting the data to the standard model for the thickness dependence of the magnetic anisotropy[4]

$$K_{\text{eff}} t_{\text{FeCoB}}^{\text{eff}} = \left( K_V - (1/2) \mu_0 M_s^2 \right) t_{\text{FeCoB}}^{\text{eff}} + K_S \qquad (1)$$

the interface and bulk anisotropy energy densities are estimated to be $K_S = 0.45 \pm 0.03$ mJ/m² and $K_V = 0.60 \pm 0.03$ MJ/m³, respectively. This value of $K_S$ is smaller than typical for Ta|FeCoB|MgO multilayers, while $K_V$ is similar to a recent report[5].

The in-plane effective demagnetization field $\mu_0 M_{\text{eff}}$ for $t_{\text{FeCoB}} = 1.6$ nm and $t_{\text{FeCoB}} = 1.8$ nm ||Ta(1)|Pt(4)|Hf($t_{\text{Hf}}$)|FeCoB($t_{\text{FeCoB}}$)|MgO(1.6)|Ru(2) samples are shown in Fig. S2(c). From these data it is clear that the insertion of a thin layer Hf at the interface of Pt and FeCoB has a significant effect on $\mu_0 M_{\text{eff}}$, with a local *minimum* at $t_{\text{Hf}} = 0.5$ nm for both series. We tentatively attribute this behavior to the role of the Hf insertion layer in both reducing the positive volume anisotropy effect from elastic strain from the underlying Pt, and in enhancing the surface anisotropy energy through reduction of strain at the FeCoB|MgO interface.

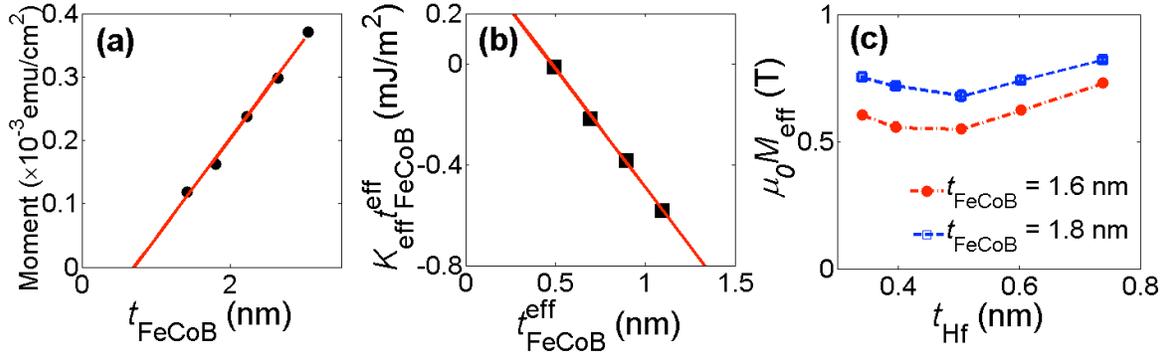

**Figure S2.** **(a)** Magnetic moment per unit area versus FeCoB thickness of $t_{Hf} = 0.5$ nm samples. From the linear fits (line), the saturation magnetization and thickness of the "dead layer" of FeCoB are $1.56 \pm 0.06 \times 10^6$ A/m and $0.7 \pm 0.1$ nm. **(b)** Magnetic anisotropy energy as a function of FeCoB effective thickness $t_{FeCoB}^{eff}$. From the linear fit (line), the interface and bulk anisotropy energy densities are $0.45 \pm 0.03$ mJ/m$^2$ and $0.60 \pm 0.03$ MJ/m$^3$, respectively. **(c)** Effective demagnetization field $\mu_0 M_{eff}$ versus Hf thickness $t_{Hf}$ of $t_{FeCoB} = 1.6$ nm (circles) and $t_{FeCoB} = 1.8$ nm samples (squares). Broken lines connect the data points.

**S4. Properties of the Pt|FeCoB control sample**

From the vibrating sample magnetometry measurement on

||Ta(1)|Pt(4)|FeCoB($t_{FeCoB}$)|MgO(1.6)|Ru(2) annealed samples, the saturation magnetization was determined to be $M_s = (1.0 \pm 0.1) \times 10^6$ A/m with no apparent magnetic dead layer. The same structure was patterned into microstrips used for frequency-dependent ST-FMR measurement from which the Gilbert magnetic damping parameter was determined, as shown in Fig. S3. As reported in the main text the best fit to the spin pumping prediction yielded $G_{eff}^{\uparrow\downarrow} = (1.1 \pm 0.1) \times 10^{15}$ $\Omega^{-1}\text{m}^{-2}$ with the damping parameter for isolated FeCoB be $0.005 \pm 0.001$ which is consistent with that obtained in Fig. 2(b) (main text).

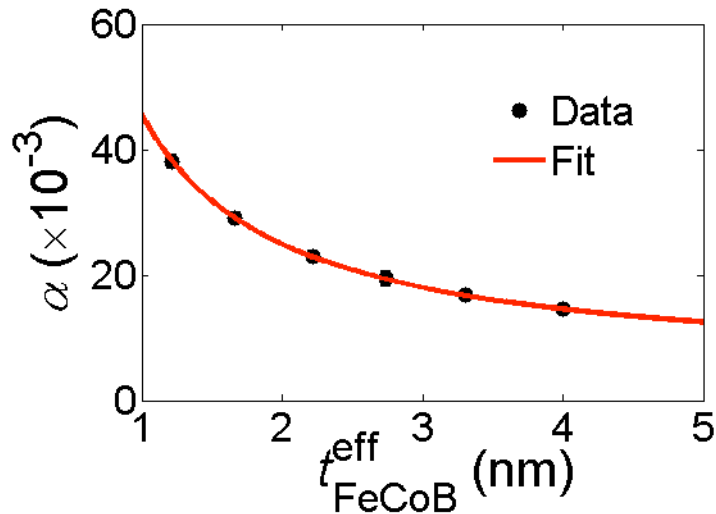

**Figure S3.** Magnetic damping parameter versus FeCoB thickness of the control samples.

**S5. Resistivity measurements**

The resistance of a film of a comparatively good metal, such as Pt, when placed between two much higher resistivity layers, such as either Ta and FeCoB, or Ta and Hf|FeCoB, begins, once the middle film's thickness is reduced to the order of twice its "bulk" elastic mean free path $l$, to be determined by diffusive scattering at the interfaces and within the adjoining layers. This effect can be modeled with the Fuchs-Sondheimer formulation[6]. Since $l_{Pt}$ in our films is ~ 2 nm, while that of the other adjoining metals are considerably shorter, interfacial scattering plays an important role in determining the electrical properties of the spin Hall structure, and also must be taken into proper consideration when considering the flow of the spin current from the bulk of the Pt layer to the Pt|Hf|FeCoB interface. From the viewpoint of the electrical energy required to effect a requisite spin transfer torque on the ferromagnetic free layer, what matters of course is the resistance per square $R_{\square,SH}$ of the spin Hall nanostrip, which we find to be $R_{\square,SH} \approx 120 \Omega/\square$ for an annealed ||Ta(1)|Pt(4)|Hf(0.5)|FeCoB(2) multilayer. This result can be alternatively expressed in terms of an averaged Pt resistivity $\bar{\rho}_{Pt} = R_{\square,SH} t_{Pt} = 50\ \mu\Omega \cdot cm$ for $t_{Pt} = 4$ nm. If a thinner Pt layer is used, $\bar{\rho}_{Pt}$ will be higher due to the interfacial scattering, shorter effective mean free path. Finally we note that there was approximately a 10% increase in $\bar{\rho}_{Pt}$ upon annealing our structures, which we take as indicative of enhanced diffusive scattering at the interface due to some additional intermixing, or perhaps the formation of a thicker PtHf bimetallic layer at the interface.

Of course when considering the flow of a net spin population generated by the SHE within the Pt to the interface $\bar{\rho}_{Pt}$ is not the proper quantity to employ in determining that spin conductance if $\bar{\rho}_{Pt}$ is determined largely by scattering close to or at the interface. Instead we measured the "differential resistivity" of the Pt as evaluated from the derivative of the conductance versus Pt thickness for $t_{Pt} > 3\,\text{nm}$ in Ta(1)|Pt(x)|Hf(0.5)|CoFeB multilayers which yielded a "bulk" resistivity $\rho_{Pt} \approx 20\,\mu\Omega\cdot\text{cm}$. Similar measurements where the Hf thickness was varied yielded a bulk Hf resistivity $\rho_{Hf} \approx 80\,\mu\Omega\cdot\text{cm}$ for $t_{Hf} \approx 2\,\text{nm}$. These were the values used in our drift-diffusion analysis of the spin pumping and spin Hall torque experimental results. Of course the fact that in these experiments the thickness of the Pt spin Hall metal is only comparable to the bulk mean free path $l_{Pt}$, together with the fact that the spin attenuation length, $\lambda_s^{Pt} \approx 1.2-1.4$ nm, as determined by previous work[7,8], is less than $l_{Pt}$ brings into question any prediction that is based on a drift-diffusion analysis.

## S6. Driff-Diffusion Analysis

### A. Spin pumping effect.

If we put aside the question raised above as to the applicability of a drift-diffusion treatment of a system where the thickness of a normal metal layer component is of of the order of its mean free path, or less in the case of the one or two atomic layer thick Hf insertion layer, and where the spin attenuation length of one of the metals is apparently less than its mean free path, we can employ the trilayer spin pumping model of Boone et al[9] to consider how an insertion layer can result in a reduction of $G_{ext}$. This drift-diffusion based model predicts

$$G_{ext} = \frac{G_{Hf}}{2}\left[\frac{G_{Hf}\coth\left(t_{Pt}/\lambda_s^{Pt}\right)+G_{Pt}\coth\left(t_{Hf}/\lambda_s^{Hf}\right)}{G_{Hf}\coth\left(t_{Pt}/\lambda_s^{Pt}\right)\coth\left(t_{Hf}/\lambda_s^{Hf}\right)+G_{Pt}}\right], \quad (S1)$$

where $G_{Hf}\equiv 1/(\rho_{Hf}\lambda_s^{Hf})$ and $G_{Pt}\equiv 1/(\rho_{Pt}\lambda_s^{Pt})$. For Pt and Hf thin films with $\rho_{Pt}=20\,\mu\Omega\cdot\text{cm}$ and $\rho_{Hf}=80\,\mu\Omega\cdot\text{cm}$, the spin attenuation lengths have been reported to be $\lambda_s^{Pt}=1.4$ nm [7,8] and $\lambda_s^{Hf}=1.5$ nm [10] respectively. A calculation using these parameters in equation (S1) indicates that that $G_{ext}$ can be only slightly reduced by a Hf insertion layer with $t_{Hf}<1$ nm. Thus the low effective spin-mixing conductance $G_{eff}^{\uparrow\downarrow}\approx 0.24\times 10^{15}\,\Omega^{-1}\text{m}^{-2}$ measured at $t_{Hf}\approx 0.5$ nm cannot be achieved without substantially reducing the bare spin-mixing conductance $G^{\uparrow\downarrow}$. However, a low $G^{\uparrow\downarrow}$ will enhance the back flow of the spin current from the Hf|FeCoF interface, lowering $\xi_{SH}$. Alternatively if we use a much different pair of parameters for the Hf layer, a higher resistivity $\rho_{Hf}$ and a shorter spin attenuation length $\lambda_s^{Hf}$ in equation (S1) this would result in a lower $G_{ext}$. However since the experimental evidence is that Hf has no significant SHE and thus that the spin current has to originate within the Pt, a shorter $\lambda_s^{Hf}$ would also result in a strong attenuation of the spin current before it reaches the Hf|FeCoB interface as discussed below.

### B. *Attenuation of the spin current in a trilayer structure*

Within the context of the drift-diffusion analysis[11] of the spin back-flow in a trilayer structure, Equation 3 in the main text is modified to be:

$$\xi_{SH}(t_{Hf}) = \theta_{SH}^{Pt} \times \frac{2\left[\cosh\left(t_{Pt}/\lambda_s^{Pt}\right)-1\right]}{\frac{\cosh(t_{Pt}/\lambda_s^{Pt})\sinh(t_{Hf}/\lambda_s^{Hf})}{\rho_{Hf}\lambda_s^{Hf}} + \frac{\cosh(t_{Hf}/\lambda_s^{Hf})\sinh(t_{Pt}/\lambda_s^{Pt})}{\rho_{Pt}\lambda_s^{Pt}}} \times G_{eff}^{\uparrow\downarrow}(t_{Hf}) \quad . \text{(S2)}$$

The rapid decrease of damping $\alpha$ that we find with our samples as a function of increasing $t_{Hf}$, as shown in Fig. 2(a) of the main text, implies the same degree of reduction in $G_{eff}^{\uparrow\downarrow}$, from Equation S2 we should expect an even more rapid reduction in $\xi_{SH}$. This expectation is not in accord with the approximately constant level of the observed $\xi_{SH}$ for $t_{Hf} \leq 0.5$ nm as shown in Fig. 4 of the main text. We conclude that the drift-diffusion analysis of the spin back-flow is simply not even approximately applicable to this spin Hall effect system with a very thin Hf insertion layer, perhaps due to dominance of interfacial processes, rather than bulk scattering, on the spin transport. As mentioned in the main text it appears that a more appropriate Boltzmann analysis of the interfacial spin transmissivity and a more detailed treatment of the electronic structure of the interfaces is likely to be required to understand this beneficial effect of the thin Hf insertion layer in enhancing the anti-damping spin torque efficiency of the Pt spin Hall effect.